\documentclass[showpacs,epsf,twocolumn]{revtex4-1}
\usepackage{float}
\usepackage{color,graphicx}
\graphicspath{ {images/} }
\usepackage{hyperref}
\begin{document}

\title{Magnetotransport properties and evidence of topological insulating state in LaSbTe}

\author{Ratnadwip Singha, Arnab Pariari, Biswarup Satpati, Prabhat Mandal}

\affiliation{Saha Institute of Nuclear Physics, HBNI, 1/AF Bidhannagar, Calcutta 700 064, India}
\date{\today}

\begin{abstract}
In this report, we present the magnetotransport and magnetization properties of LaSbTe single crystals. Magnetic field-induced turn-on behavior and low-temperature resistivity plateau have been observed. By adopting both metal-semiconductor crossover and Kohler scaling analysis, we have discussed the possible origin of the temperature and magnetic field dependence of resistivity. At 5 K and 9 T, a large, non-saturating transverse magnetoresistance (MR) $\sim$ 5$\times$10$^{3}$ \% has been obtained. The MR shows considerable anisotropy, when the magnetic field is applied along different crystallographic directions. The non-linear field dependence of the Hall resistivity confirms the presence of two types of charge carriers. From the semiclassical two-band fitting of Hall conductivity and longitudinal conductivity, very high carrier mobilities and almost equal electron and hole densities have been deduced, which result in large MR. The Fermi surface properties have been analyzed from de Haas-van Alphen oscillation. From the magnetization measurement, the signature of non-trivial surface state has been detected, which confirms that LaSbTe is a topological insulator, consistent with the earlier first-principles calculations.
\end{abstract}

\maketitle

\section{Introduction}

The complex electronic band structure in condensed matter systems always fascinates researchers as it leads to unique states of matter, often reflected in their unusual electronic transport properties. In this regard, the recent discovery of topological insulator (TI) and three-dimensional topological semimetal (TSM) has emerged as a major boost, which has unfolded a whole new domain to explore the dynamics of relativistic particle in low-energy electronic system. The experimental discovery of TI \cite{Chen,Xia,Zhang} was followed by the realization of three-dimensional TSMs, which attracted enormous attention due to the presence of relativistic Dirac and Weyl fermions as quasi-particle excitations in their symmetry protected semimetallic bulk state. Among the TSM candidates, Cd$_{3}$As$_{2}$ and Na$_{3}$Bi were first theoretically predicted \cite{Wang1,Wang2} and experimentally verified \cite{Liu1,Liu2} to host Dirac fermions, whereas Weyl fermions were identified in $TX$ ($T$ = Ta, Nb; $X$ = As, P) family of materials \cite{Weng,Lv,Lv2,Xu1,Huang,Xu2,Xu3,Yang,Sekhar}. In addition to the fundamental physics, the unique transport responses such as extreme magnetoresistance (XMR), ultrahigh electronic mobility, make these materials highly preferable for technological applications. Therefore, prediction and experimental finding of new materials with topological non-trivial electronic band structure are of growing interest.\\

Recently, from band structure calculations, Xu \textit{et al.} \cite{Xu4} have proposed that the members of the family \textit{WHM} (\textit{W} being Zr, Hf or La; \textit{H} is from group IV or V; \textit{M} is group VI element) are potential candidates for two-dimensional (2D) TI. Angle-resolved photoemission spectroscopy (ARPES) has revealed 2D topological insulating state on the surface of ZrSnTe \cite{Lou}, whereas several other members, ZrSiS, ZrSiSe and ZrSiTe have been confirmed as topological nodal-line semimetals from ARPES and transport experiments \cite{Schoop,Singha,Ali,Hu1}. Therefore, detailed studies on different members of this family are required, which may lead to novel quantum states of matter.

In this report, we present the magnetotransport and magnetization properties of single crystalline LaSbTe, a member of the \textit{WHM} family. Our experimental results reveal compensated electron-hole density with very high carrier mobility in LaSbTe. The observed magnetic field-induced resistivity upturn and large MR are analyzed both from the viewpoints of possible metal-semiconductor-like transition and Kohler scaling. The Fermi surface parameters have been calculated from de Haas-van Alphen oscillation in the magnetization measurement. In the low-field region of the magnetization measurement data, a robust paramagnetic singularity has been observed, which originates from the spin-polarized non-trivial surface state. Our results thus confirm a topological insulating phase in LaSbTe, which is in accordance with the theoretical prediction \cite{Xu4}.

\section{Sample preparation and experimental details}

The single crystals of LaSbTe were grown by molten-salt flux method \cite{DiMasi}. A mixture of LiCl (Alfa Aesar 99.9\%) and RbCl (Alfa Aesar 99.8\%) in 55:45 molar ratio was used as flux. The chloride salt mixture along with stoichiometric amount of La (Alfa Aesar 99.9\%), Sb (Alfa Aesar 99.9999\%) and Te (Alfa Aesar 99.999\%) were taken in an alumina crucible which was then sealed in a quartz tube under vacuum. The quartz tube was heated to 700$^{\circ}$C and kept at this temperature for 5 days. After that, the furnace was cooled slowly (2$^{\circ}$C/h) to room temperature. Shinny single crystals were obtained which were washed with water to remove the chloride salt and cleaned with acetone. X-ray diffraction (XRD) patterns were obtained in a Rigaku X-ray diffractometer (TTRAX III). High resolution transmission electron microscopy (HRTEM) of the crystals was done in FEI, TECNAI G$^{2}$ F30, S - TWIN microscope operating at 300 kV and equipped with GATAN Orius SC1000B CCD camera. Transport measurements were performed in a 9 T physical property measurement system (Quantum Design) with ac transport option on several single crystal samples from the same batch, all shaped into thin rectangular bar geometry. Magnetic measurements were done in a 7 T SQUID-VSM MPMS 3 (Quantum Design). Before doing the magnetic measurements on LaSbTe samples, we have measured the empty sample holders to ensure the absence of any contamination. The obtained data are more than two orders of magnitude smaller than that obtained with LaSbTe. Similar properties have been observed for all the crystals.

\section{Results and discussions}

\subsection{Sample characterization}
In the inset of Fig. 1(a), a single crystal of LaSbTe of typical dimensions 2 mm$\times$1 mm$\times$0.6 mm is shown with different crystallographic axes along which measurements have been done. The XRD pattern of powdered single crystals  is illustrated in Fig. 1(a). The sharp diffraction peaks with small full-width at half-maximum confirm the high quality of the grown crystals. The XRD pattern has been analyzed by Rietveld structural refinement using FULLPROF software package and the peaks have been indexed. LaSbTe crystallizes in a ZrSiS type structure and belongs to orthorhombic space group \textit{Pmcn} \cite{DiMasi,Wang3}. The refined lattice parameters calculated from XRD spectra, \textbf{a}=4.3903(2), \textbf{b}=4.4293(3) and \textbf{c}=19.4858(3) \AA, are consistent with previous report \cite{DiMasi}. The selective area electron diffraction (SAED) pattern obtained in HRTEM is shown in Fig. 1(b) with the Miller indices of the corresponding lattice planes.

\subsection{Temperature dependence of resistivity and low-temperature resistivity plateau.}

In Fig. 2(a), the resistivity of two samples from the same batch is shown as a function of temperature. Both the samples exhibit identical temperature dependence, decreasing almost linearly from 300 K down to 125 K and show weak temperature dependence below 30 K. The resistivity at 2 K is as low as $\sim$9 $\mu$$\Omega$ cm for sample 1 and $\sim$6 $\mu$$\Omega$ cm for sample 2, which are comparable to that reported for Dirac semimetal Cd$_{3}$As$_{2}$ and Weyl semimetal candidates TaAs, NbP and TaP \cite{Sekhar,Huang,Hu2,Xiang}. The calculated residual resistivity ratio [RRR=$\rho_{xx}$(300 K)/$\rho_{xx}$(2 K)] for both the samples (sample 1 $\sim$8.4; sample 2 $\sim$14), suggests good metallicity of the grown LaSbTe crystals. Although these values of RRR are significantly smaller than that observed in ZrSiS (RRR $\sim$288)\cite{Singha}, it is comparable to that reported for another sister compound HfSiS \cite{Kumar}. From the structural point of view, which also plays an important role on electronic properties, HfSiS (tetragonal) is rather closer to ZrSiS than the present one (orthorhombic; LaSbTe). Therefore, it may not be very straight forward to compare the RRR of LaSbTe with ZrSiS. We would also like to mention that the RRR in Dirac/Weyl semimetals has been seen to vary over a wide range. The observed value of RRR for LaSbTe is, however, comparable to several topological semimetals \cite{Huang,Hu2,Xiang}. For both the crystals, the low-temperature resistivity can be fitted well [Fig. 2(a) inset] with $\rho_{xx}(T)=A + BT^{n}$ type relation for \textit{n}$\sim$3, where \textit{A} and \textit{B} are constants. Deviation from a value \textit{n}=2 is generally considered as a departure from the pure electronic correlation dominated scattering mechanism \cite{Ziman}. Similar type of temperature dependence of resistivity has also been observed in unconventional semimetals LaSb (\textit{n}=4) and LaBi (\textit{n}=3), elemental yttrium  and transition metal carbide and has been attributed to interband electron-phonon scattering \cite{Tafti,Sun,Hall,Zhang2}.\\

\begin{figure}
\includegraphics[width=0.45\textwidth]{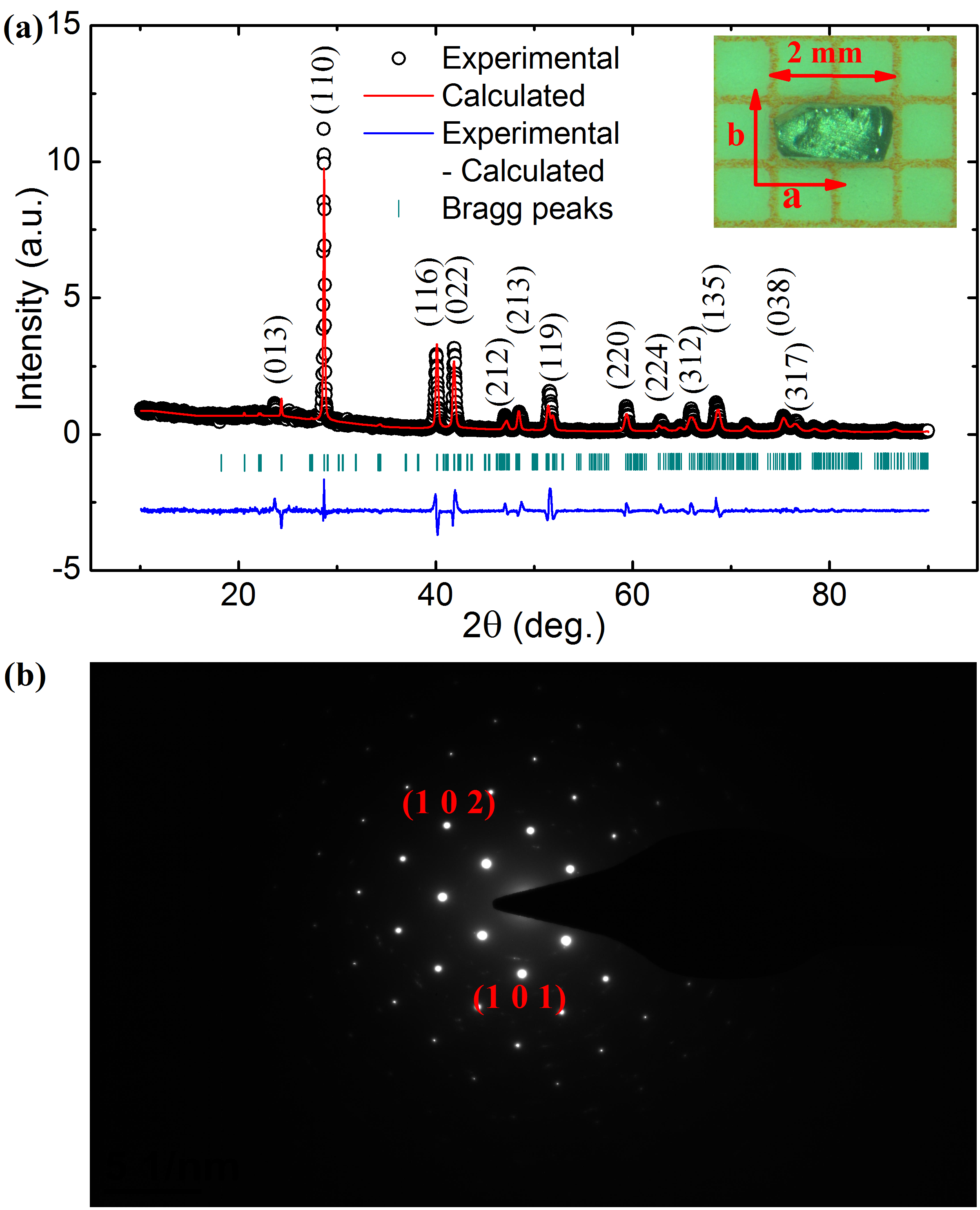}
\caption{(Color online) (a) Powder XRD pattern of the LaSbTe single crystals. Inset shows a typical single crystal. (b) SAED pattern obtained in HRTEM measurement.}
\end{figure}

\begin{figure}
\includegraphics[width=0.35\textwidth]{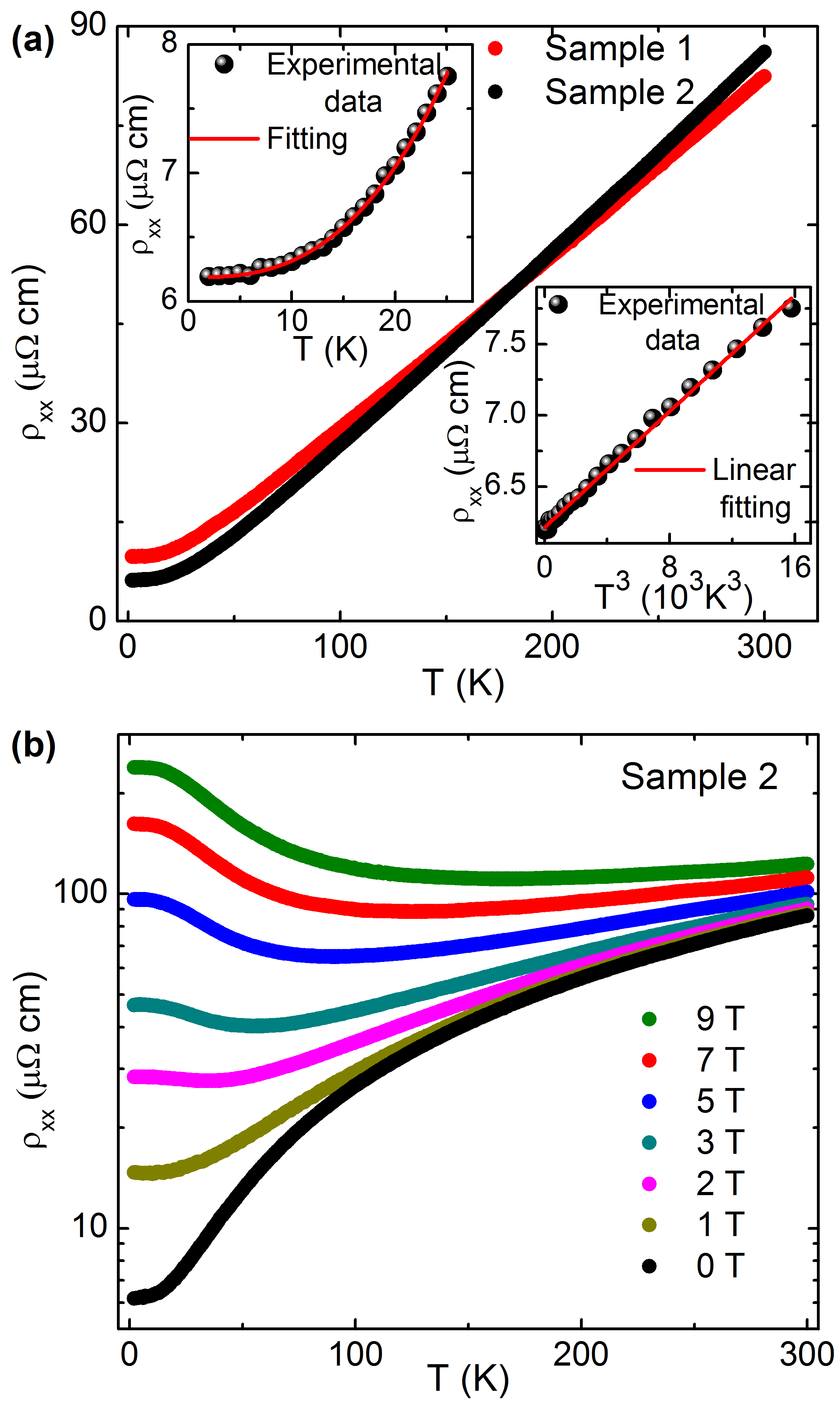}
\caption{(Color online) (a) Temperature dependence of resistivity for two samples from the same batch. Upper inset shows the low-temperature region fitted with $\rho_{xx}(T)=A + BT^{n}$. The low-temperature resistivity is plotted with T$^{3}$ in the lower inset. (b) $\rho_{xx}(T)$ under different transverse magnetic field strengths.}
\end{figure}

When a magnetic field is applied, resistivity is observed to increase rapidly, particularly in the low-temperature region [Fig. 2(b)]. Above a critical value of the magnetic field $\sim$1 T, the resistive behavior of LaSbTe at low temperature modifies significantly. The nature of the slope of the $\rho_{xx}$($T$) curve changes and a metal-semiconductor-like crossover appears. Similar upturn in low-temperature resistivity has been observed in several TSMs \cite{Sekhar,Hu2,Tafti,Sun}. In analogy to dynamical chiral symmetry breaking in the relativistic theory of (2 + 1)-dimensional Dirac fermions, Khveshchenko proposed that this type of magnetic field-induced crossover is due to the gap opening at the band crossing points  \cite{Khveshchenko}. However, subsequent explanations in terms of scaling analysis, have also been proposed \cite{Du,Kopelevich,Wang4}. Here, we have adopted both the approaches to discuss the resistivity turn-on behavior in LaSbTe. $\rho_{xx}$($T$) can be analyzed considering the thermal activated transport, $\rho_{xx}($T$)$$\propto$exp(E$_{g}$/k$_{B}T$), as in the case of intrinsic semiconductor and several topological semimetal systems \cite{Tafti,Wang5}. In Fig. 3(a), we have plotted  $\ln$$\rho_{xx}$ as a function of $T^{-1}$. From the slope of the curve, the thermal activation energy gap (E$_{g}$) $\sim$2.6 meV has been calculated at 9 T, which is quite small. As shown in Fig. 3(b), the calculated energy gap clearly shows a strong magnetic field dependence.\\

\begin{figure}
\includegraphics[width=0.5\textwidth]{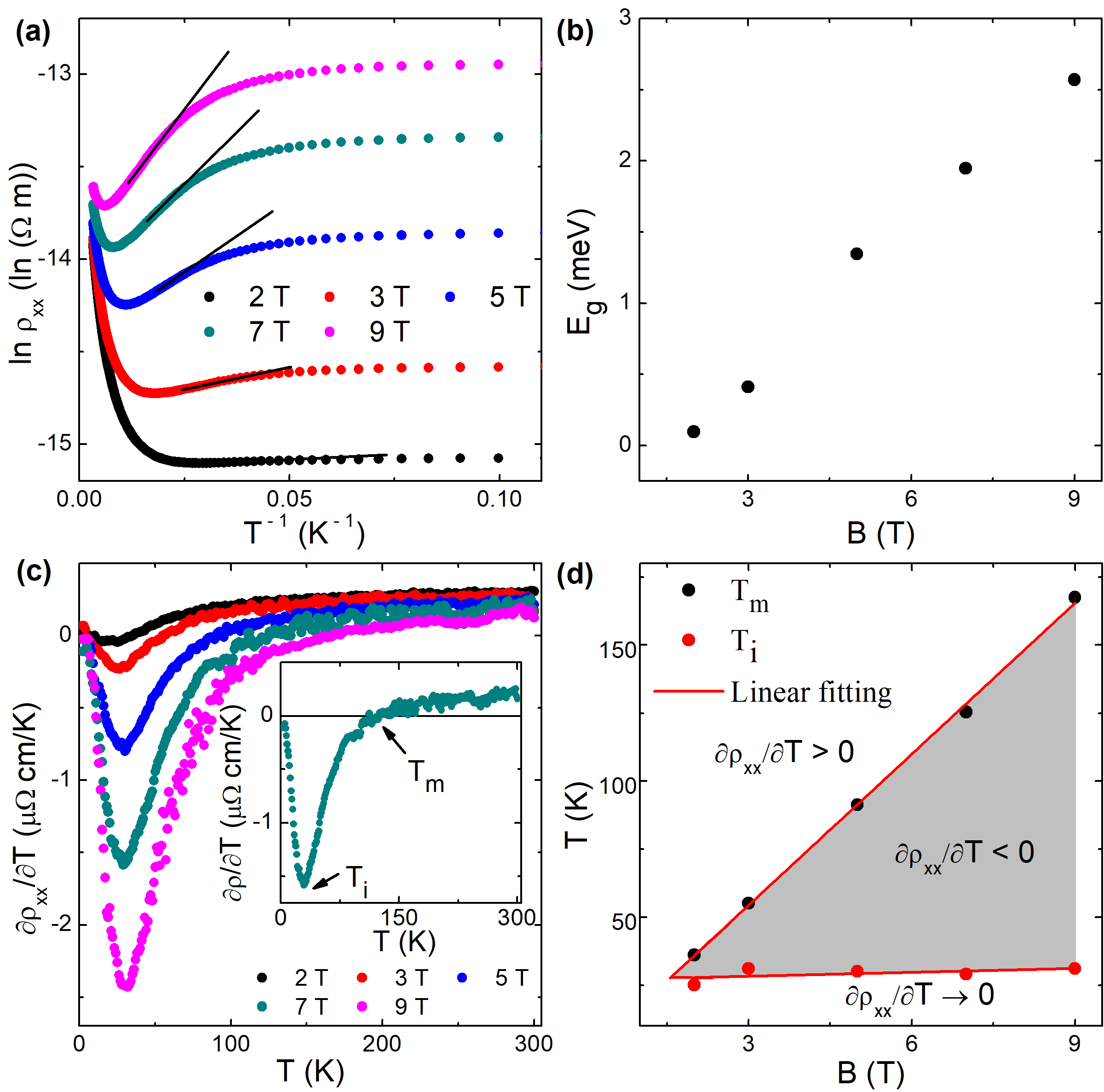}
\caption{(Color online) (a) $\ln$$\rho$ plotted as a function of $T^{-1}$. (b) Field dependence of the calculated thermal activation energy gap. (c) $\partial\rho_{xx}$/$\partial T$ plotted as a function of temperature for different applied field strengths. Inset shows the crossover temperature ($T_{m}$) and plateau temperature ($T_{i}$) for 7 T. (d) Triangular temperature-field phase diagram constructed from $T_{m}$ and $T_{i}$.}
\end{figure}

Above the critical magnetic field, the crossover in $\rho_{xx}$($T$) curve is followed by a plateau at low temperature. Although, the field-induced resistivity plateau has been observed in several topologically non-trivial semimetallic systems \cite{Sekhar,Hu2,Tafti,Sun,Wang4}, its origin is not yet settled unambiguously. Very similar temperature-dependent resistivity has been observed in topological insulators Bi$_{2}$Te$_{2}$Se and SmB$_{6}$  in absence of magnetic field, where the low-temperature saturation-like behavior in resistivity arises due to the competition between conducting surface and insulating bulk states \cite{Ren,Kim1}. On the other hand, the field-induced resistivity plateau in LaSbTe appears in broken time-reversal symmetry scenario. In Fig. 3(c), the first-order derivative of the resistivity $\partial\rho_{xx}$/$\partial T$ is plotted as a function of temperature for different magnetic field strengths. As shown in the inset, from the resultant curves, two distinct characteristic temperatures can be identified, the crossover temperature $T_{m}$, where $\partial\rho_{xx}$/$\partial T$ changes sign and $\emph{T}_{i}$, below which the resistivity plateau starts to appear. While $T_{m}$  increases monotonically with  field, $T_{i}$ is almost field independent. In Fig. 3(d), following the approach of Tafti \textit{et al.} \cite{Tafti1}, we have constructed a triangular temperature-field phase diagram for LaSbTe by plotting $T_{m}$ and $T_{i}$ as a function of magnetic field. This triangular phase diagram has been seen to be universal for all semimetallic systems showing extreme magnetoresistance \cite{Tafti1}. In the figure, the gray shaded area which corresponds to $\partial\rho_{xx}$/$\partial T$$<$0, denotes the region where XMR occurs. The linear fitted curves of $T_{m}$($B$) and $T_{i}$($B$) merge at B$_{0}$$\sim$1.6 T, which is precisely the turn-on field above which metal-semiconductor-like crossover appears. The region above the shaded triangle with $\partial\rho_{xx}$/$\partial T$$>$0 is of metallic conduction and negligible MR, whereas the area below the shaded triangle denotes the plateau region ($\partial\rho_{xx}$/$\partial T$$\rightarrow$0). As has been seen in bismuth and graphite \cite{Du}, the triangular region in the temperature-field phase diagram can be specified by the inequality, $\hbar/\tau \lesssim \hbar\omega_{c} \lesssim k_{B}T$, where $\tau$ is the electron-phonon scattering time and $\omega_{c}$ is the cyclotron frequency. In clean semimetals with low carrier density, $\tau$$^{-1}$$\ll$k$_{B}$T/$\hbar$ and hence, there exists a wide temperature-field range where XMR and  metal-semiconductor-like crossover in resistivity appear. On the other hand, large carrier density and strong impurity scattering, limit the MR in conventional metals \cite{Abrikosov}.

\begin{figure}
\includegraphics[width=0.5\textwidth]{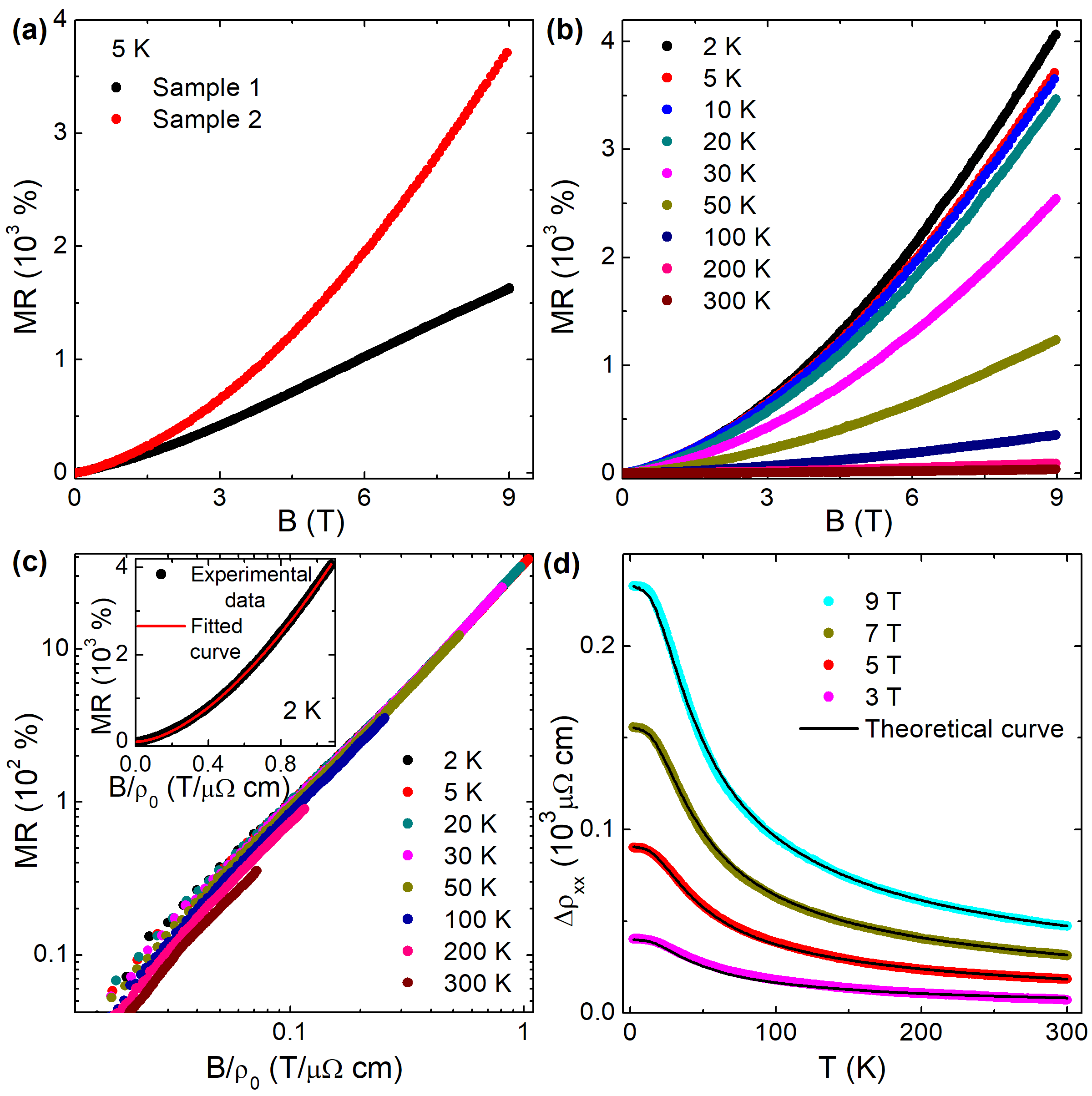}
\caption{(Color online) (a) Transverse MR for sample 1 and sample 2. (b) Transverse MR for sample 2 at different representative temperatures. (c) MR at different temperatures are scaled using Kohler's rule. Inset shows the power-law (Eq. 1) fitting of the experimental data. (d) Field-dependent part of $\rho_{xx}$(T). Solid lines denote the generated curves from Kohler's rule.}
\end{figure}

\subsection{Large and anisotropic magnetoresistance in LaSbTe.}

Next, we have measured the transverse MR, i.e., MR under transverse electric and magnetic fields. In Fig. 4(a), the MR for two samples is compared at 5 K with current along \textbf{a} and field along \textbf{c} axis. It is clear from figure that the sample with higher RRR (sample 2) shows stronger response to magnetic field than the other one (sample 1). Therefore, the value of MR clearly depends on the RRR. However, the MR for both the samples is observed to follow same power-law behavior, MR$\propto$B$^{m}$ with \textit{m} $\sim$1.6. In Fig. 4(b), we have plotted  MR at different temperatures for sample 2 as a representative. At 2 K and 9 T, a large MR $\sim$4$\times$10$^{3}$ \% is obtained without any signature of saturation. Though the observed MR is not among the largest reported so far, it is comparable to that observed in several topological semimetals \cite{Xiang,Novak,Wakeham,Pavlokiuk}. With the increase in temperature, however, the MR decreases rapidly to only $\sim$35 \% at 300 K and 9 T. As shown in Fig. 4(c), employing the Kohler's rule,
\begin{equation}
MR=\alpha (B/\rho_{0})^{m},
\end{equation}
with $\alpha$=4.8$\times$10$^{-9}$ ($\Omega$ cm/T)$^{1.6}$ and \textit{m}=1.6, the MR curves at different temperatures can be scaled to a single curve. According to semiclassical two-band theory, the validity of Kohler rule with MR$\propto$(B/$\rho_{0}$)$^{2}$ suggests a perfectly compensated system \cite{Ziman}. In LaSbTe, however, the exponent is not exactly 2 and as a result a small deviation from scaling in Fig. 4(c) is observed at high temperature. The deviation becomes more prominent for temperature above 100 K. Several aspects \cite{Wang4,McKenzie}, including different densities and or temperature dependence of mobility for two types of carriers may lead to the violation of Kohler scaling in LaSbTe, to be discussed later on. Adopting a modified form of Eq. (1),
\begin{equation}
\rho_{xx}(B)=\rho_{0} + \alpha B^{m}/\rho_{0}^{m-1},
\end{equation}
it can be shown that the resistivity in a magnetic field consists of two components, zero-field resistivity $\rho$$_{0}$ and field-induced component $\Delta\rho_{xx}$=$\alpha B^{m}/\rho_{0}^{m-1}$. The competition between these two terms may give rise the observed minimum in the temperature dependence of resistivity above the critical field. In Fig. 4(d), the field-induced resistivity component is plotted for different magnetic fields. As illustrated by the solid lines, the experimental data at all magnetic fields can be described well by Eq. (2) with $\alpha$=4.8$\times$10$^{-9}$ ($\Omega$ cm/T)$^{1.6}$ and \textit{m}=1.6.  MR at different magnetic fields in fact collapses onto a single curve when normalized by their respective values at 2 K [Fig. 5(a)]. This suggests that the temperature dependence of MR remains same for all magnetic fields. On the other hand, the calculated thermal activation energy gap (E$_{g}$) has been observed to increase monotonically with field. So, sharper metal-semiconductor-like crossover is expected at higher magnetic fields. Therefore, the validity of such scaling behavior, seems to be contradictory to the picture of possible field-induced gap at the band crossing points. Moreover, the gap opening model has several shortcomings. For example, the calculated small energy gap ($\sim$2.6 meV) corresponds to a temperature $\sim$25 K. However, the semiconductor-like behavior persists up to a temperature as high as $\sim$150 K. In addition to that, no change in Hall resistivity has been observed through out the measured temperature range. As $\alpha$ and \textit{m} are temperature-independent constants, Eq. (2) suggests that the temperature dependence of $\Delta\rho_{xx}$ is determined entirely by $\rho$$_{0}$($T$), which is again inversely proportional to density and mobility of the charge carrier. Therefore, in a system where carrier density remains constant with temperature, the resistivity turn-on behavior may result from strong temperature dependence of the carrier mobility \cite{Wang4,Du}. In spite of being conflicting, both metal-semiconductor crossover \cite{Tafti,Wang5} and Kohler scaling analysis \cite{Du,Kopelevich,Wang4} are used simultaneously and can explain the observed transport behavior to some extent. However, the actual origin remains an open question, which is yet to be answered decisively.\\

\begin{figure}
\includegraphics[width=0.5\textwidth]{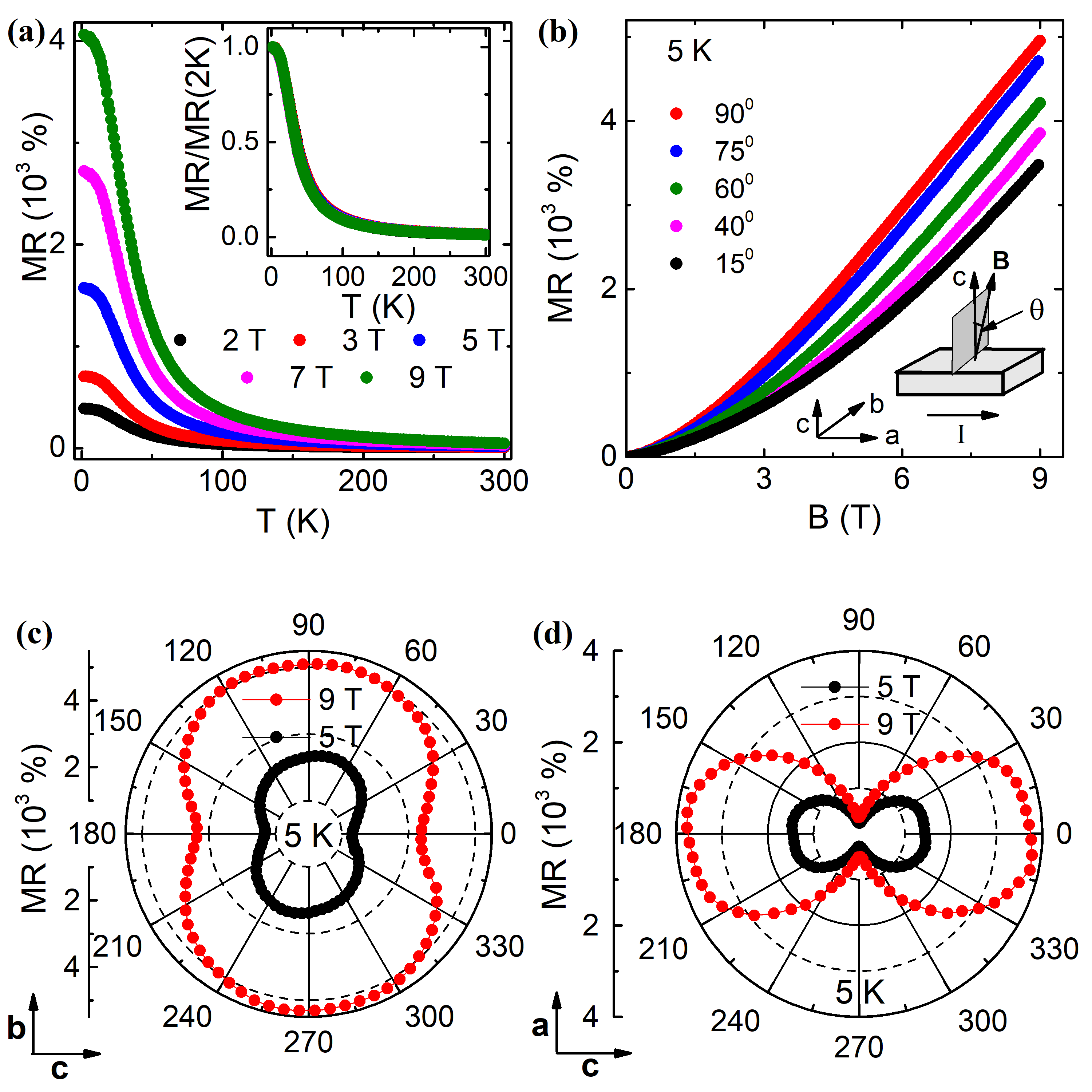}
\caption{(Color online) (a) Temperature dependence of MR. Inset shows the normalized MR with the value at 2 K. (b) MR as a function of magnetic field at 5 K when B is tilted along different crystallographic directions. Inset shows the schematic of the experimental setup. (c) Polar plot of TMR for two different field strengths with current along \textbf{a} axis and magnetic filed rotated in the \textbf{bc} plane. (d) Angular dependence of MR with current parallel to \textbf{a}-axis and field rotated in the \textbf{ac} plane.}
\end{figure}

For in-depth understanding of magnetotransport properties of LaSbTe, we have measured the MR by tilting the magnetic field along different crystallographic directions while keeping the current direction unaltered. The experimental configuration is shown schematically in the inset of Fig. 5(b) with current along \textbf{a} axis and magnetic field is rotated within the \textbf{bc} plane. As we increase the angle ($\theta$) from 0$^{\circ}$ to 90$^{\circ}$, MR is seen to increase and then decreases with further increase of $\theta$ beyond 90$^{\circ}$. Maximum MR $\sim$5$\times$10$^{3}$ is recorded at $\theta$ $\sim$90$^{\circ}$ (\textbf{B}$\parallel$\textbf{b}). In Fig. 5(c), MR for two different field strengths, is plotted as a function of the tilting angle. The resultant curve possesses a two-fold rotational symmetry, indicating considerable anisotropy in the magnetotransport properties in LaSbTe. Anisotropic MR has also been reported in other members of the \textit{WHM} family \cite{Singha,Hu1,Ali}. The MR in ZrSiS and ZrSiSe shows strong anisotropy and buttery-like angular dependence. \cite{Hu1,Ali}. On the other hand, the MR in ZrSiTe exhibits a two-fold symmetric pattern, which is similar to the present system \cite{Hu1}.

We have also measured the MR with current along \textbf{a} axis and magnetic field is rotated within the \textbf{ac} plane. As shown in the Fig. 5(d),  MR becomes minimum when electric and magnetic fields are parallel to each other ($\varphi$=90$^{\circ}$; \textbf{B}$\parallel$\textbf{a}), which is expected due to the absence of Lorentz force in such a configuration. MR increases monotonically as the angle decreases from 90$^{\circ}$ to 0$^{\circ}$. For both the angle variation measurement configurations, weak but detectable kinks have been observed symmetrically at certain angles (at 60$^{0}$ and 240$^{0}$ for field rotated in \textbf{bc} plane; at 165$^{0}$ and 345$^{0}$ for field rotated in \textbf{ac} plane). It may be due to some higher order texturing, which has also been reported in another isostructural compound ZrSiS \cite{Ali}.

\subsection{Hall measurement.}

\begin{figure}
\includegraphics[width=0.5\textwidth]{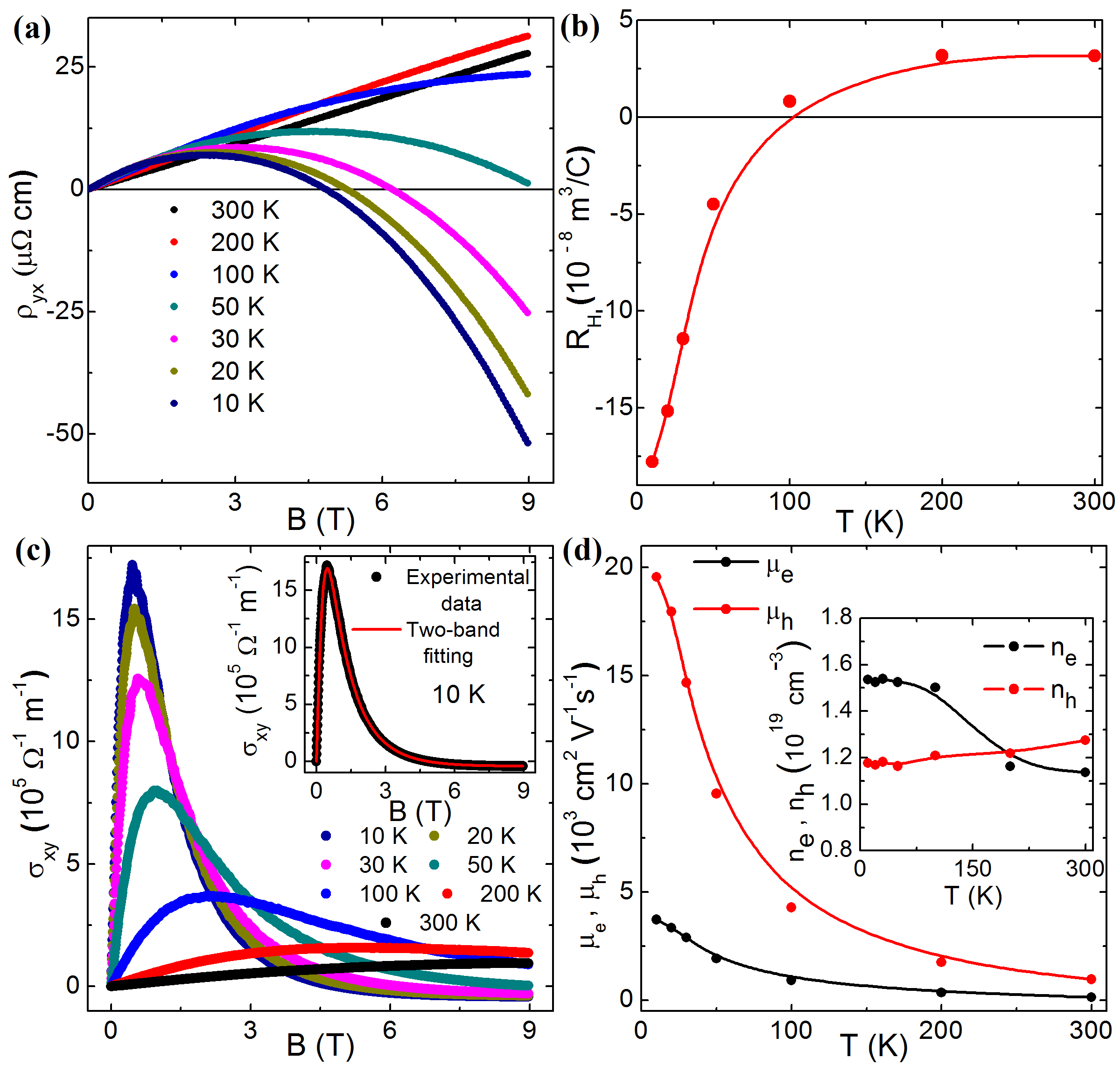}
\caption{(Color online) (a) Field dependence of the Hall resistivity at different temperatures. (b) Temperature dependence of the Hall coefficient obtained from the slopes of $\rho$$_{yx}(B)$ curves in high-field region. At about 100 K, the Hall coefficient changes sign from positive to negative. (c) The Hall conductivity $\sigma$$_{xy}$=$\frac{\rho_{yx}}{\rho_{yx}^{2} + \rho_{xx}^{2}}$, has been obtained from the tensorial inversion of the resistivity matrix and plotted as a function magnetic field. Inset shows the two-band fitting of the Hall conductivity at 10 K using Eq. (3). (d) Temperature dependence of electron and hole carrier mobility obtained from two-band fitting. Inset shows the extracted carrier densities as a function of temperature.}
\end{figure}

To determine the nature and density of the charge carriers, we have performed the Hall resistivity measurement in the temperature range 10-300 K. In Fig. 6(a), the measured Hall resistivity ($\rho$$_{yx}$) is plotted as a function of magnetic field. At 300 K, $\rho$$_{yx}$ is found to be almost linear and positive which indicate hole-dominated charge conduction. However, with decreasing temperature, $\rho$$_{yx}$($B$) becomes sublinear and, around 100 K, its sign changes from positive to negative at high fields. The magnetic field dependence of $\rho$$_{yx}$ clearly shows that  more than one type of charge carrier is present in LaSbTe. From the high-field slope of the $\rho$$_{yx}$($B$) curves, the Hall coefficient has been determined and shown in Fig. 6(b) as a function of temperature. Following the semiclassical two-band model \cite{Huang,Hurd}, we have fitted the Hall conductivity $\sigma$$_{xy}$ in Fig. 6(c) using,
\begin{equation}
\sigma_{xy}=\left[n_{h}\mu_{h}^{2}\frac{1}{1+(\mu_{h}B)^2}-n_{e}\mu_{e}^{2}\frac{1}{1+(\mu_{e}B)^2}\right]eB,
\end{equation}
where $\sigma$$_{xy}$=$\frac{\rho_{yx}}{\rho_{yx}^{2} + \rho_{xx}^{2}}$ has been obtained from the tensorial inversion of the resistivity matrix. $\rho$$_{xx}$ is the longitudinal resistivity. $n_{e}$ ($n_{h}$) and $\mu$$_{e}$ ($\mu$$_{h}$) are electron (hole) density and mobility, respectively. From the fitting parameters, the electron and hole densities at 10 K are found to be, 1.5(3)$\times$10$^{19}$ cm$^{-3}$ and 1.2(2)$\times$10$^{19}$ cm$^{-3}$ respectively, which indicate near-perfect electron-hole carrier compensation in LaSbTe. The obtained electron and hole mobilities are large and at 10 K are $\sim$3.7(1)$\times$10$^{3}$ cm$^{2}$ V$^{-1}$ s$^{-1}$ and $\sim$1.9(2)$\times$10$^{4}$ cm$^{2}$ V$^{-1}$ s$^{-1}$ respectively, which are comparable to the carrier mobility observed in Dirac semimetals Cd$_{3}$As$_{2}$, ZrTe$_{5}$, TlBiSSe and Weyl semimetal WTe$_{2}$ \cite{Novak,Narayanan,Zheng,Luo}. As shown in Fig. 6(d), the mobility for both types of carriers decreases as the temperature increases. The inset of Fig. 6(d) illustrates the temperature dependence of the extracted carrier density for both electrons and holes. While the carrier densities remain almost same throughout the temperature range, the mobility shows strong temperature dependence and it is different for the two types of charge carriers. Due to the uncertainties in the numerical values of the parameters obtained from the two-band fitting, Fig. 6(d) provides only a qualitative information about two types of carriers. At temperature above $\sim$100 K, the mobilities for electrons and holes become comparable and both types of carriers start to contribute significantly to the transport properties. This behavior is reflected in the Kohler scaling analysis, where the MR curves have been seen to deviate from the scaling at higher temperature.

To verify the results obtained from Hall resistivity, we have also analyzed the longitudinal conductivity ($\sigma_{xx}=\frac{\rho_{xx}}{\rho_{xx}^{2}+\rho_{yx}^{2}}$) using two-band model \cite{Huang,Zhang5},
\begin{equation}
\sigma_{xx}=e\left[n_{h}\mu_{h}\frac{1}{1+(\mu_{h}B)^{2}}+n_{e}\mu_{e}\frac{1}{1+(\mu_{e}B)^{2}}\right].
\end{equation}
In Fig. 7, we have shown the global fitting of $\sigma$$_{xy}$ and $\sigma_{xx}$, i.e., these two quantities have been fitted simultaneously using Eq. (3) and (4) to enhance the reliability of the extracted parameters. The obtained electron [hole] density $\sim$2.5(3)$\times$10$^{19}$ cm$^{-3}$ [$\sim$1.9(5)$\times$10$^{19}$ cm$^{-3}$] and mobility $\sim$3.3(1)$\times$10$^{3}$ cm$^{2}$ V$^{-1}$ s$^{-1}$ [$\sim$1.2(8)$\times$10$^{4}$ cm$^{2}$ V$^{-1}$ s$^{-1}$] at 10 K, are in good agreement with that calculated from the Hall measurements.

\begin{figure}
\includegraphics[width=0.35\textwidth]{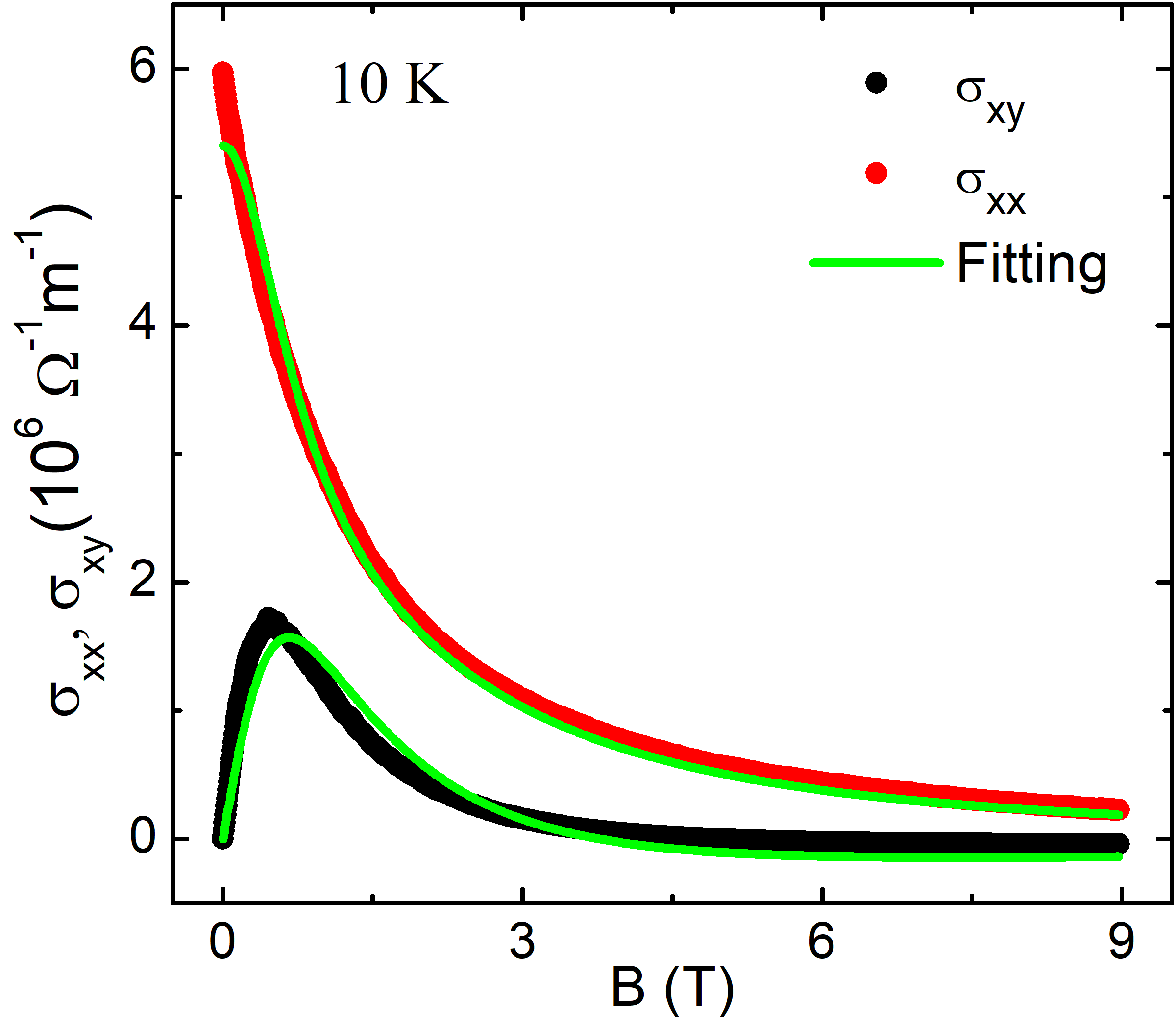}
\caption{(Color online) Global fitting of the Hall conductivity and longitudinal conductivity ($\sigma_{xx}=\frac{\rho_{xx}}{\rho_{xx}^{2}+\rho_{yx}^{2}}$) using two-band model [Eq. (3) and (4)].}
\end{figure}

\subsection{de Haas-van Alphen oscillation and Fermi surface topology.}

\begin{figure}
\includegraphics[width=0.5\textwidth]{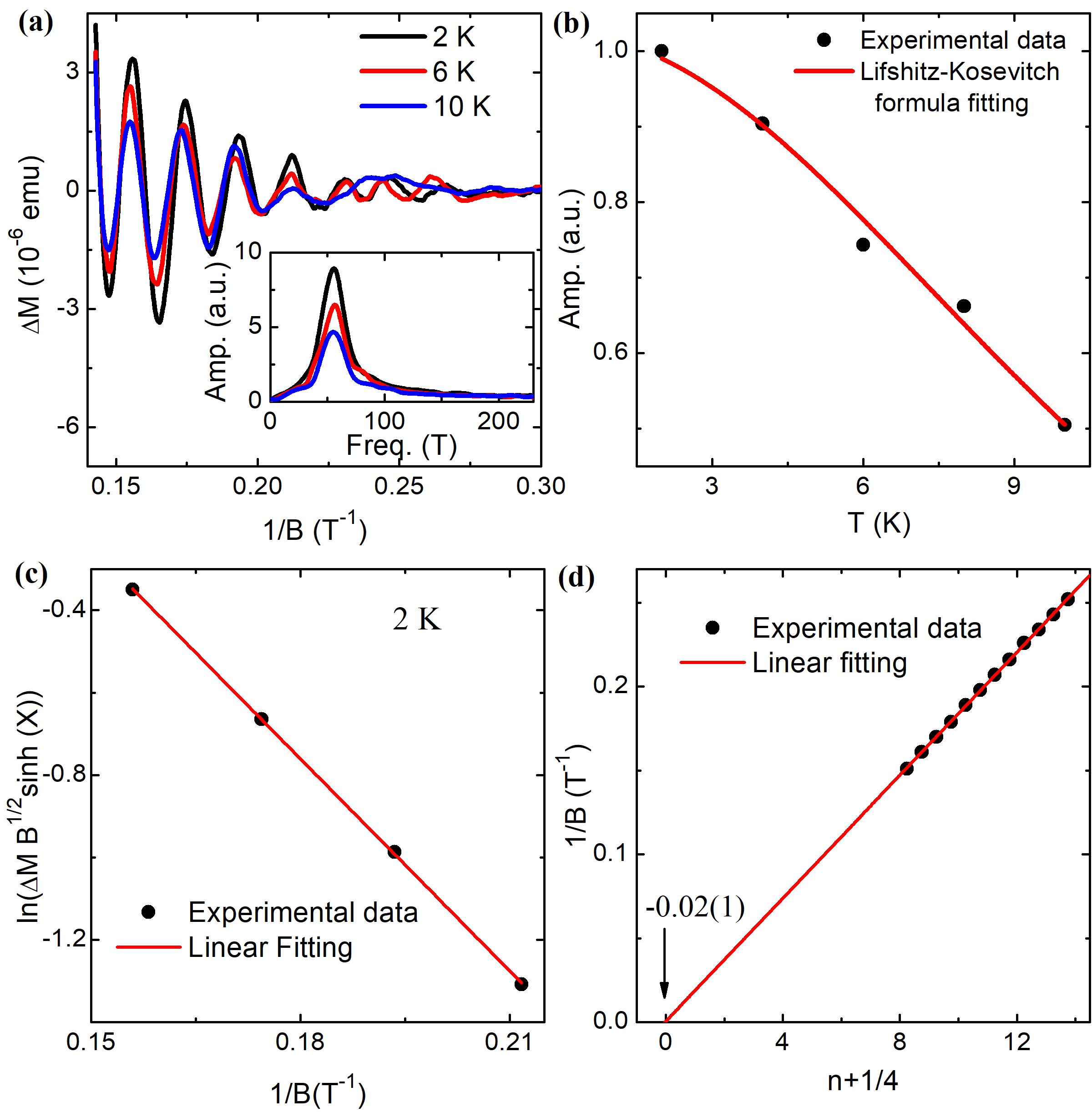}
\caption{(Color online) (a) dHvA oscillation in the magnetization data with field along \textbf{c}-axis, obtained after background subtraction. Inset shows the corresponding FFT result. (b) Temperature dependence of the oscillation amplitude, fitted using the thermal damping term of Lifshitz-Kosevich (L-K) formula. (c) Field dependence of the oscillation amplitude at 2 K. Solid line is the fit to the experimental data using L-K formula, where $X=2\pi^{2}k_{B}Tm^{\ast}/e\hbar B$. (d) Landau level fan diagram with maxima position as $n$+1/4 and minima position as $n$+3/4. The arrow indicates the x-axis intercept.}
\end{figure}

We have performed the magnetization measurement on the LaSbTe single crystals with magnetic field along \textbf{c}-axis. The field dependence of the magnetization data show a diamagnetic behavior with prominent de Haas-van Alphen (dHvA) oscillation up to 10 K. Above this temperature, the thermal scattering of the carriers starts to dominate and the oscillations quickly washed out. To extract the oscillatory component ($\Delta M$), we have subtracted a smooth background from the magnetization data and plotted in Fig. 8(a) as a function of $1/B$. The corresponding fast Fourier transform spectrum in the inset, shows a single frequency $\sim$55(1) T. From the Hall resistivity data, it is evident that there are one electron-type and one hole-type Fermi pockets in LaSbTe. However, as the density of charge carrier for both the Fermi pockets is almost equal, their volume is expected to be equivalent. In addition, the weak anisotropy in the angular dependence of MR indicates nearly spherical geometry of the Fermi pockets. So, the cross-sectional areas of the Fermi pockets must be almost equal and as a consequence, it is difficult to distinguish them through quantum oscillation measurements. Similar behavior has also been observed for Dirac semimetal Cd$_{3}$As$_{2}$, where two equivalent ellipsoidal Fermi pockets result in single frequency quantum oscillation \cite{He}. To ensure that this quantum oscillation corresponds to the bulk Fermi pocket, we have also done the magnetization measurements along crystallographic \textbf{a} and \textbf{b}-axis. For both directions, dHvA oscillations have been observed. The corresponding FFT spectra (Appendix Fig. 10) show almost similar oscillation frequency [75(3) T for \textbf{B}$\parallel$\textbf{a}-axis and 77(2) T for \textbf{B}$\parallel$\textbf{b}-axis] along three mutually perpendicular directions, indicating the three-dimensional nature of the Fermi surface in LaSbTe. Moreover, the estimated carrier density ($\sim$5$\times$10$^{18}$ cm$^{-3}$) from the oscillation frequency is close to that obtained from Hall measurement and is orders of magnitude higher than that expected ($\sim$10$^{16}$) for any two-dimensional surface state \cite{Analytis}. Using the Onsager relation, $F=(\phi_{0}/2\pi^{2})A_{F}$, where $\phi_{0}$ is the single magnetic flux quantum, we have calculated the Fermi surface cross-section ($A_{F}$) $\sim$5.2(1)$\times$10$^{-3}$ ${\AA}^{-2}$ perpendicular to the \textbf{c}-axis. The corresponding Fermi momentum ($k_{F}$) is $\sim$4.0(1)$\times$10$^{-2}$ ${\AA}^{-1}$. The temperature dependence of the oscillation amplitude is plotted in Fig. 8(b) and has been fitted using the thermal damping factor of Lifshitz-Kosevich (L-K) formula, $R_{T} = (2\pi^{2}k_{B}T/\beta)/sinh(2\pi^{2}k_{B}T/\beta)$, where $\beta = e\hbar B/m^{\ast}$. $m^{\ast}$ is the cyclotron mass of the carriers and estimated to be 0.06(2)$m_{0}$ from the fitting parameters, where $m_{0}$ is the rest mass of a free electron. In Fig. 8(c), the magnetic field dependence of the oscillation amplitude at 2 K is shown. The experimental data have been fitted using the magnetic field dependent part of the L-K formula, $\Delta M = -B^{1/2} R_{T} exp(-2\pi^{2}k_{B}m^{\ast}T_{D}/\hbar eB)$, where $T_{D}$ is the Dingle temperature. From the fitting parameters, Dingle temperature, quantum mobility [$\mu_{q}$=$(e\hbar/2\pi k_{B}m^{\ast}T_{D})$] and the mean free path ($l$) of the charge carriers have been estimated. All the extracted parameters from the quantum oscillation are listed in Table I. The calculated quantum mobility of the charge carriers is seen to be smaller than the classical Drude mobility, which has been obtained from the Hall resistance. This is expected as $\mu_{q}$ is sensitive to both large- and small-angle scattering in contrast to classical mobility, which is only influenced by large-angle scattering \cite{Narayanan}.

\begin{table*}[t]
\begin{center}
\textbf{TABLE I: Fermi surface parameters extracted from dHvA oscillation.}

\begin{tabular}{c c c c c c c c c c c c c c c}\\
\hline
F & $A_{F}$ & $k_{F}$ & $m^{\ast}$ & $v_{F}$ & $T_{D}$ & $\mu_{q}$ & $l$\\

T & 10$^{-3} {\AA}^{-2}$ & 10$^{-2} {\AA}^{-1}$ & $m_{0}$ & 10$^{5}$ m/s & K & 10$^{3}$cm$^{2}$V$^{-1}$s$^{-1}$ & nm\\ [0.5ex]
\hline\hline

55(1) & 5.2(1) & 4.0(1) & 0.06(2) & 8(3) & 19(2) & 1.8(3) & 50(26)\\[0.5ex]
\hline
\end{tabular}
\end{center}
\end{table*}

Further information about the topological nature of the band structure in LaSbTe, can be obtained from the Berry phase associated to the motion of the charge carrier. For the parabolic band dispersion the Berry phase takes a value 0, whereas a non-trivial $\pi$ Berry phase is obtained for linearly dispersing bands. An additional phase contribution ($\delta$) appears with a value 0 or $\pm$1/8 for two-dimensional and three-dimensional band structures, respectively. The Berry phase can be determined from the x-axis intercept of a Landau level fan diagram, which is plotted from the maxima and minima positions of the quantum oscillation. Considering the peak positions of $\Delta$$M$ as integer+1/4 and valley positions as half integer+1/4, the x-axis intercept will be 0 for trivial electronic band and 0.5 for non-trivial band structure \cite{Hu1,Shoenberg,Sergelius}. In Fig. 8(d), we have plotted the Landau level fan diagram from the dHvA oscillation. The obtained intercept $\sim$-0.02(1) is close to the theoretical value for trivial band structure and far from that expected (0.5) for a non-trivial state. Hence, the observed Berry phase confirms that the bulk band structure of LaSbTe is topologically trivial.

\subsection{Signature of the helical spin texture in the surface state.}

\begin{figure}
\includegraphics[width=0.5\textwidth]{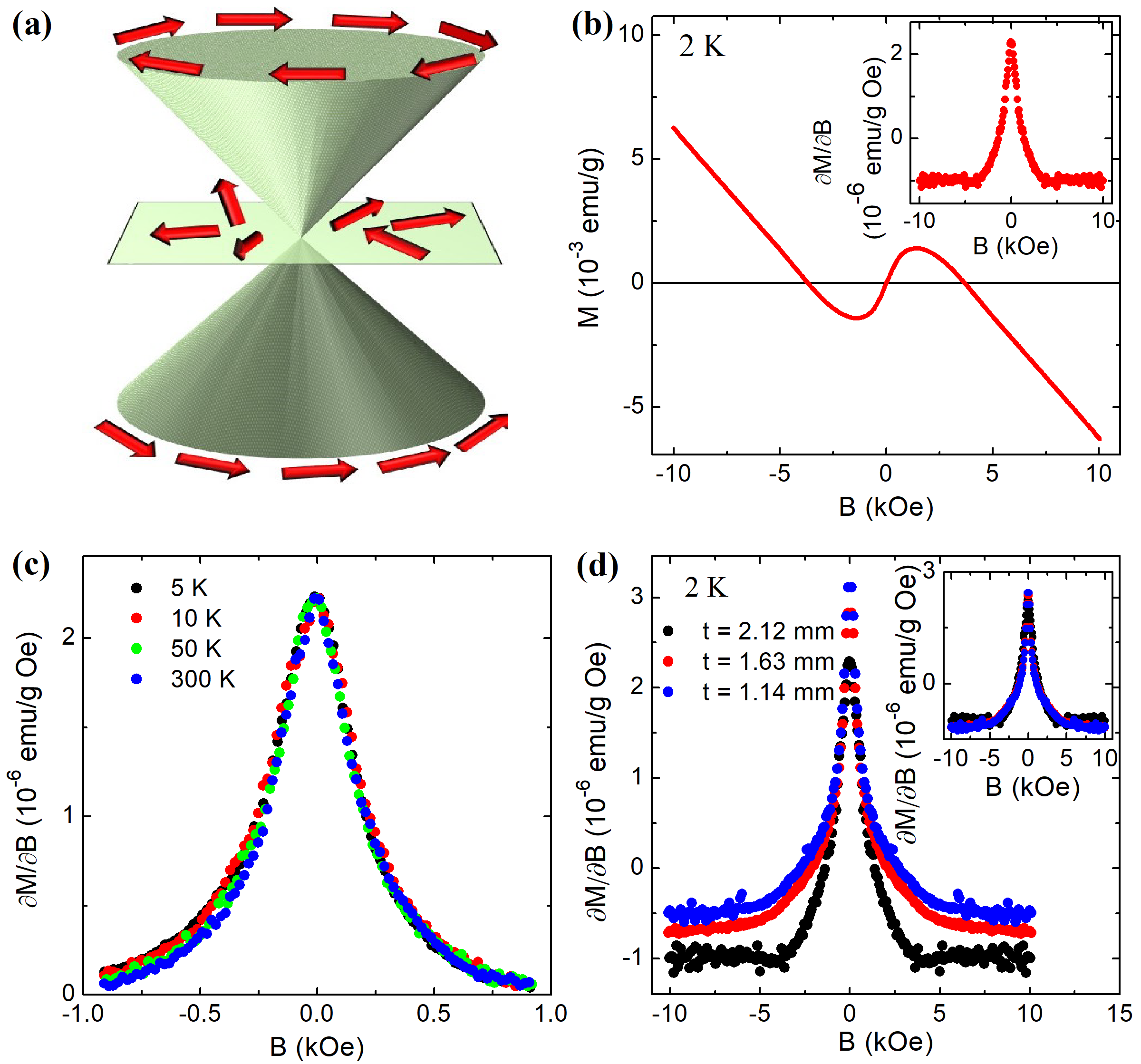}
\caption{(Color online) (a) Schematic representing the spin-texture of the surface state in a 3D topological insulator. The arrows indicate the direction of the electron spins. (b) Low field region of the magnetization data at 2 K for LaSbTe. Insets shows the corresponding magnetic susceptibility $\chi$ (=$\partial M/\partial B$). (c) Magnetic susceptibility at different temperatures, indicating the robustness of the signal. (d) Magnetic susceptibility curves at 2 K for different sample thickness (t). Inset shows the normalized curves.}
\end{figure}

A TI is characterized by topologically distinct surface and bulk electronic states. While the bulk state is insulating with gap between conduction and valence band, surface state hosts linear band crossings. The surface band structure is protected by time reversal symmetry and can be represented by a Dirac-type effective Hamiltonian \cite{Zhang3}, $H_{surf}(k_{x}, k_{y}) = \hbar v_{F}(\sigma^{x}k_{y}-\sigma^{y}k_{x})$, where $\vec{\sigma}$ is the Pauli matrix. This leads to the spin-momentum locking and absence of backscattering of the charge carriers during conduction through surface channel \cite{Zhang3,Burkov}. For a fixed momentum (\textbf{k}), the electron `spin' (i.e. total angular momentum) has a fixed direction, which forms a spin-texture \cite{Hsieh,Pan}. The electron-spin texture can be described in terms of a helicity operator $\hat{h} = (1/k)\hat{z}.(\vec{k}\times\vec{\sigma})$, which takes values +1 and -1 for the lower and upper Dirac cones of the surface electronic band structure, respectively \cite{Zhang3}. As a consequence of opposite spin helicity for the upper and lower Dirac cones, the electronic states close to the Dirac point face singularity in their spin orientation, provided the Dirac spectrum is not gapped. These small number of electrons do not have any preferable spin alignment and the spins are randomly oriented. An external magnetic field can align the spins along its direction and as a result a paramagnetic contribution is observed in the total magnetic moment \cite{Zhao2}. The corresponding magnetic susceptibility shows a cusp in $\chi(B)$ plot. The spin-texture of the surface state in a 3D topological insulator is shown schematically in Fig. 9(a). In Fig. 9(b), the low-field region of the magnetization curve is shown for LaSbTe, at a representative temperature 2 K. Although, diamagnetic behavior is observed at higher field, a clear paramagnetic signal appears in the low-filed region. The susceptibility (shown in the inset) exhibits a cusp at $B=0$, as expected for a topological insulator \cite{Zhao2}. Similar field dependence of magnetic susceptibility has been reported for several topological insulators such as Bi$_{2}$Se$_{3}$, Bi$_{2}$Te$_{3}$, Sb$_{2}$Te$_{3}$, Bi$_{1.5}$Sb$_{0.5}$Te$_{1.7}$Se$_{1.3}$ \cite{Zhao2,Buga,Dutta} and narrow gap topological semimetals ZrTe$_{5}$ and LaBi \cite{Pariari,Singha2}, both with spin helical Dirac cone surface states. With temperature and chemical potential are set to zero, the total susceptibility can be mathematically formulated as \cite{Zhao2},
\begin{equation}
\chi(B) \cong \chi_{0} + \frac{\mu_{0}}{4\pi^{2}}\frac{x}{L}\left[\frac{(g\mu_{B})^{2}}{\hbar v_{F}}\Lambda - \frac{2(g\mu_{B})^{3}}{\hbar^{2}v_{F}^{2}}\mid B\mid\right].
\end{equation}
Here $\chi_{0}$, $\Lambda$, $\mu_{B}$, $g$ and $L$ are the background contribution, effective size of the momentum space contributing to the singular part of the total free energy, Bohr magneton, Land$\acute{e}$ $g$-factor and thickness of the measured crystal, respectively. $x$ is the fraction of the surface state contributing to the areal susceptibility. A linear field decay of $\chi(B)$ is expected from Eq. (5), which is indeed observed for LaSbTe single crystals. Another interesting feature of such magnetic response is the `robustness' of the paramagnetic signal. As shown in Fig. 9(c), the susceptibility cusp exists even at room temperature and the peak height along with its sharpness are invariant of the temperature. This unusual thermal stability of the paramagnetic signal can be attributed to an intrinsic surface cooling process of thermoelectric origin \cite{Zhao2}. Such unique temperature independent behavior of susceptibility can not be explained assuming a small paramagnetic or ferromagnetic impurity in diamagnetic host and is completely different from the properties of dilute magnetic semiconductors. Moreover, the standard diamagnetic (Bi), paramagnetic (Pd) samples (Fig. 11 in the Appendix) and topological Dirac semimetal Cd$_{3}$As$_{2}$, do not show such phenomena \cite{Pariari}. To confirm the origin of the paramagnetic singularity, we have done the magnetization measurements by changing the sample thickness (t). The corresponding susceptibility curves are shown in Fig. 9(d). With decreasing bulk volume, the diamagnetic background has been seen to decrease and approximately scales with the thickness of the crystal. However, as shown in the inset, the height of the paramagnetic response remains unaltered, similar to that observed in Bi$_{2}$Se$_{3}$, Bi$_{2}$Te$_{3}$, Sb$_{2}$Te$_{3}$ \cite{Zhao2}. Therefore, the observed paramagnetic singularity must be originated from the non-trivial surface bands near the band crossing points and confirms the TI state in LaSbTe.

\section{Discussions}

LaSbTe represents a large family of isostructural compounds. The members of this family have similar electronic band structure and proposed to be weak TIs from theoretical calculations \cite{Xu4}. The single layer of these compounds is an ideal candidate to realize a 2D TI with global band gap, which is induced by spin-orbit coupling. They lead to the weak TI, when the layers are stacked on top of another, thus forming a three-dimensional structure. From ARPES experiments, ZrSnTe (a member of \textit{WHM} family) has been confirmed to host TI state on the surface \cite{Lou}. On the other hand, several members of this family are shown to be topological nodal-line semimetals \cite{Schoop,Singha,Hu1}. So, it is equally intriguing to explore the topological nature of other members of the family. The magnetotransport results of LaSbTe are identical to several TSMs. However, investigations on the surface state reveal a TI band structure. These results are somewhat similar to that obtained for ZrTe$_{5}$ and LaBi \cite{Pariari,Singha2}. While both of the compounds show semimetallic transport properties \cite{Sun,Li}, from surface probing ARPES experiments and magnetic measurements \cite{Zhang4,Nayak,Pariari,Singha2}, these materials are identified as TIs with narrow gap in the bulk band structure. In these systems, the semimetallic properties appear due to large bulk carrier density, which suppresses the surface state contribution.

\section{Conclusions}

In conclusion, we present the systematic study of magnetotransport and magnetic properties of single crystalline LaSbTe. Magnetic field-induced resistivity turn-on along with low-temperature resistivity plateau have been observed and analyzed from the aspects of possible metal-semiconductor crossover as well as the Kohler's scaling. From the $\rho_{xx}(T)$ curves, a triangular temperature-field phase diagram has been constructed, which is universal for semimetals showing XMR. At 2 K and 9 T, a large transverse MR $\sim$5$\times$10$^{3}$ \% has been observed without any signature of saturation. By rotating the magnetic field along different crystallographic directions, significant anisotropy in the magnetotransport properties has been observed. From the Hall measurement, the presence of two types of carriers has been confirmed. The semiclassical two-band fitting of the Hall and longitudinal conductivity reveals near-perfect carrier compensation with very high carrier mobilities and explains the large MR in electron-hole resonance regime. The Fermi surface parameters have been calculated from de Haas-van Alphen oscillation in the magnetization measurement. In the low-field region of the magnetization data, a robust paramagnetic singularity has been detected, which is a clear signature of the helical spin-texture of the non-trivial surface state in a topological insulator. Thus our measurements unambiguously confirm a TI state in LaSbTe, which is in accordance with the theoretical calculations.

\section{Acknowledgements}

We acknowledge and thank S. Pakhira, G. Bhattacharjee and S. Roy for their help during measurements and fruitful discussions.

\begin{center}
\textbf{APPENDIX}\\

\textbf{A. dHvA oscillation along crystallographic a and b-axis.}
\end{center}

dHvA oscillation has been also observed with magnetic field along crystallographic \textbf{a} and \textbf{b}-axis. In Fig. 10, the corresponding FFT spectra are shown at 2 K. The obtained frequencies are 75(3) T for \textbf{B}$\parallel$\textbf{a}-axis and 77(2) T for \textbf{B}$\parallel$\textbf{b}-axis.
\begin{figure}[h]
\includegraphics[width=0.45\textwidth]{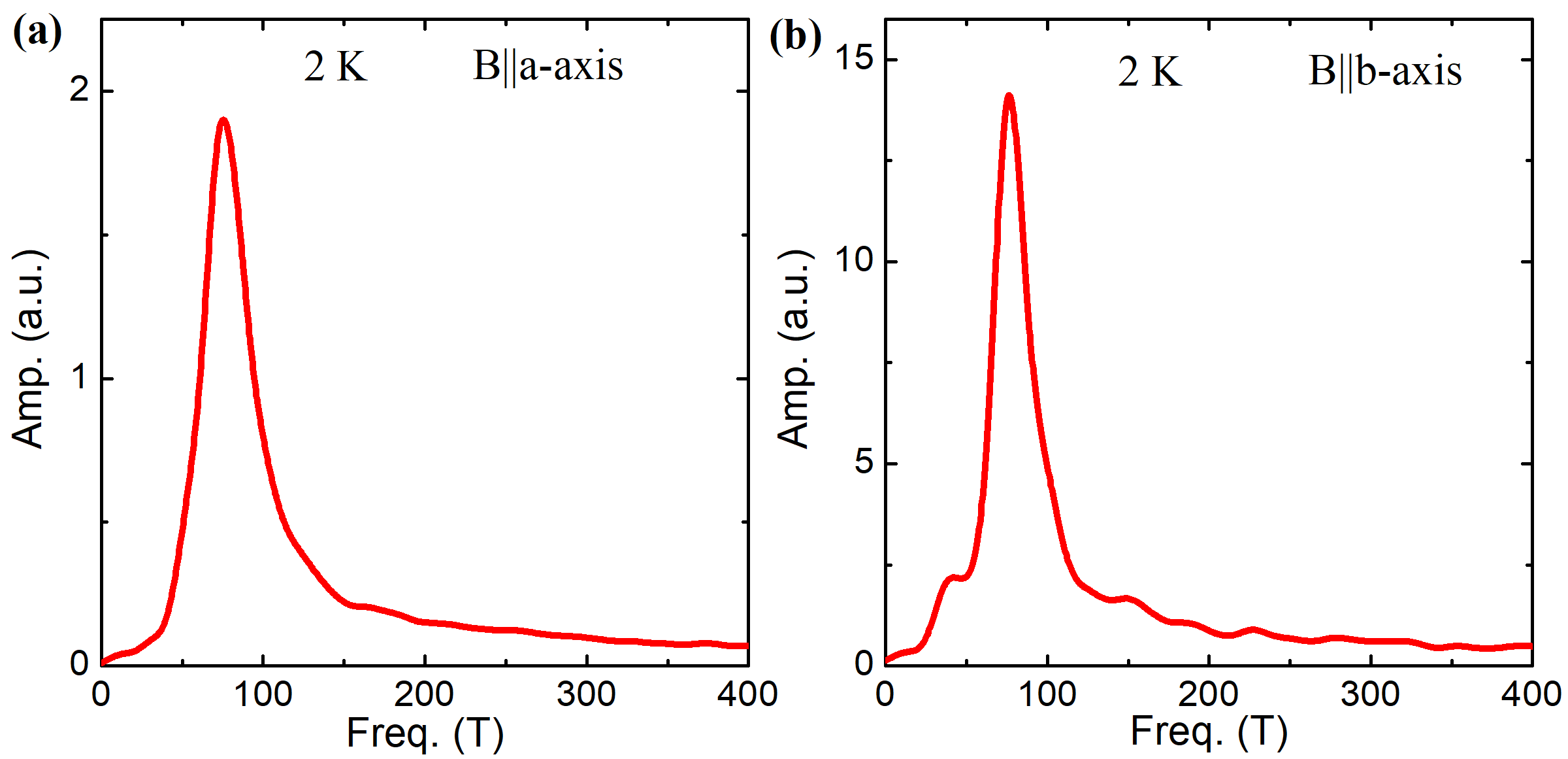}
\caption{(Color online) FFT spectra of the dHvA oscillation for magnetic field applied along \textbf{a} and \textbf{b}-axis.}
\end{figure}

\begin{center}
\textbf{B. Magnetization measurement for standard samples.}
\end{center}

The results of the magnetization measurement on standard diamagnetic (Bi) and paramagnetic (Pd) samples are shown in Fig. 10. In both the cases, no cusp-like behavior has been observed in the magnetic susceptibility data in the vicinity of $B=0$.

\begin{figure}[h]
\includegraphics[width=0.5\textwidth]{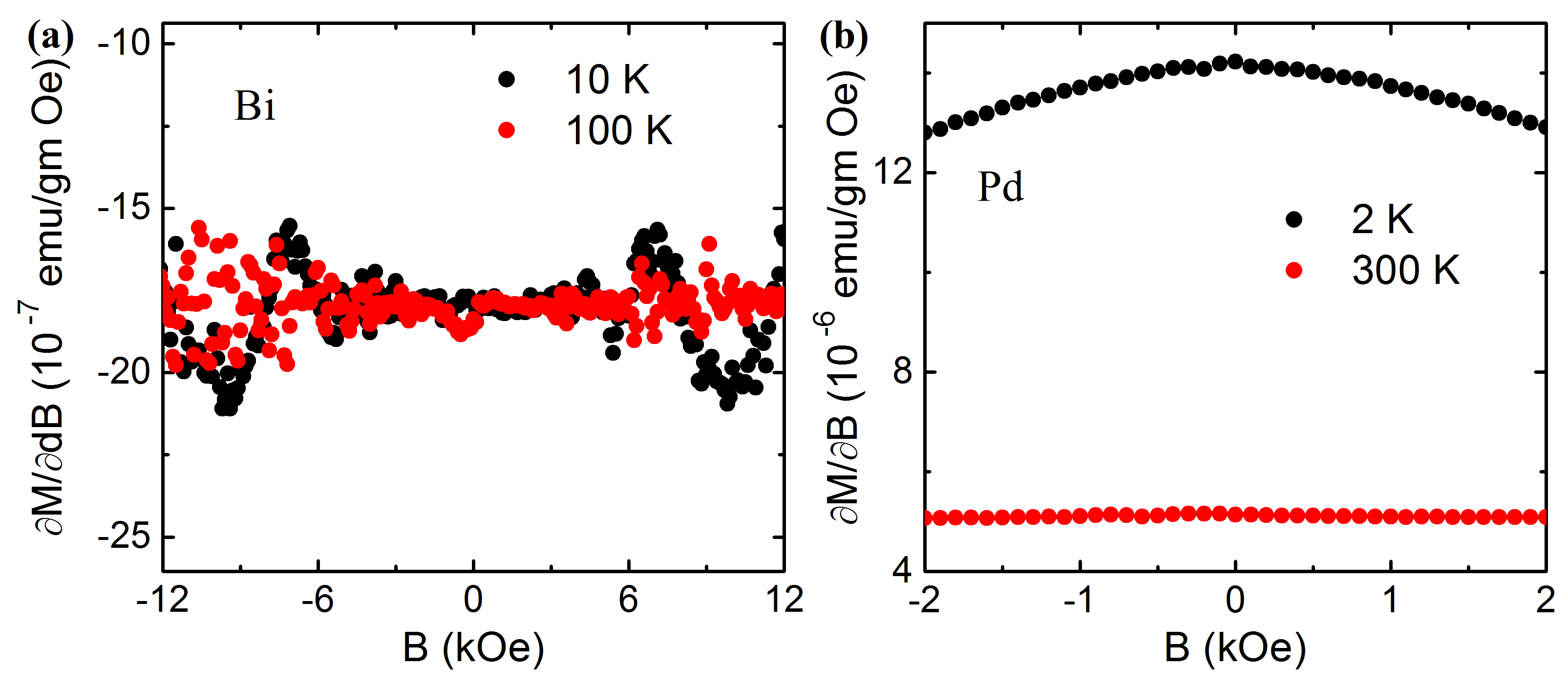}
\caption{(Color online) Magnetic susceptibility $\chi$ (=$\partial M/\partial B$) of (a) bismuth and (b) palladium at different representative temperatures.}
\end{figure}

\end{document}